\documentclass[sigconf]{acmart}

\usepackage{booktabs} 

\setcopyright{rightsretained}

\renewcommand\footnotetextcopyrightpermission[1]{}

\acmDOI{}

\acmISBN{}

\copyrightyear{2026}
\acmYear{2026}
\acmConference[PCSC2026]{Philippine Computing Science Congress}{April 2026}{Davao, Philippines}
\acmBooktitle{Proceedings of the 26th Philippine Computing Science Congress (PCSC 2026), April 23--25, 2026, Davao, Philippines}

\begin{document}
\title[Galaw at Gunita]{Galaw at Gunita: Extended Reality Murals\\ for Experiencing Filipino Art}

\author{Jomar {Delos Reyes}}
\authornote{All authors contributed equally to this research. You may contact us using firstname\_lastname@dlsu.edu.ph. }
 \affiliation{%
  \institution{De La Salle University}
  \city{Manila}
  \country{Philippines}
  }

\author{Sealtiel {Dy}} 
\authornotemark[1]
 \affiliation{%
  \institution{De La Salle University}
  \city{Manila}
  \country{Philippines}
  }

\author{Rica Mae {Sales}} 
\authornotemark[1]
 \affiliation{%
  \institution{De La Salle University}
  \city{Manila}
  \country{Philippines}
  }

\author{Orrin Landon {Uy}} 
\authornotemark[1]
 \affiliation{%
  \institution{De La Salle University}
  \city{Manila}
  \country{Philippines}
  }

\author{Toni-Jan Keith {Monserrat}} 
 \affiliation{%
  \institution{De La Salle University}
  \city{Manila}
  \country{Philippines}
  }

\author{Ryan Austin {Fernandez}} 
 \affiliation{%
  \institution{De La Salle University}
  \city{Manila}
  \country{Philippines}
  }

\author{Jordan Aiko {Deja}} 
\orcid{0000-0001-9341-6088}
 \affiliation{%
  \institution{De La Salle University}
  \city{Manila}
  \country{Philippines}
  }
\email{jordan.deja@dlsu.edu.ph}

\renewcommand{\shortauthors}{Delos Reyes, Dy, Sales, Uy et al.}

\begin{abstract}
Digital and interactive spaces are becoming increasingly prevalent as platforms for cultural engagement, offering new ways to make art more accessible, engaging, and inclusive. In the Philippine context, where visual art is deeply rooted in precolonial, colonial, and postcolonial histories, there is a growing need to explore how digital systems can support art appreciation without replacing or compromising traditional and physical artworks. Rather than treating digital experiences as substitutes, we argue for the value of creating interactive digital twins that allow audiences to explore, touch, and engage with artworks while preserving the integrity of the originals. In this paper, we present \textit{SoulWall}, an extended reality (XR) interactive mural system designed to augment Filipino artworks through embodied interaction. SoulWall enables viewers to experience paintings and animations at scale, supporting exploratory and playful engagement while respecting artist intent. We describe the design and implementation of the system, including its mural layout, interaction techniques, and interaction logging infrastructure. We report findings from a user study focused on user experience complemented by analyses of interaction logs and spatial engagement patterns. Our results highlight the potential of XR murals as a cultural computing artifact for art appreciation and for showcasing Filipino artists in interactive public and exhibition spaces.
\end{abstract}

%
%



\begin{teaserfigure}
\centering
  \includegraphics[width=0.8\linewidth]{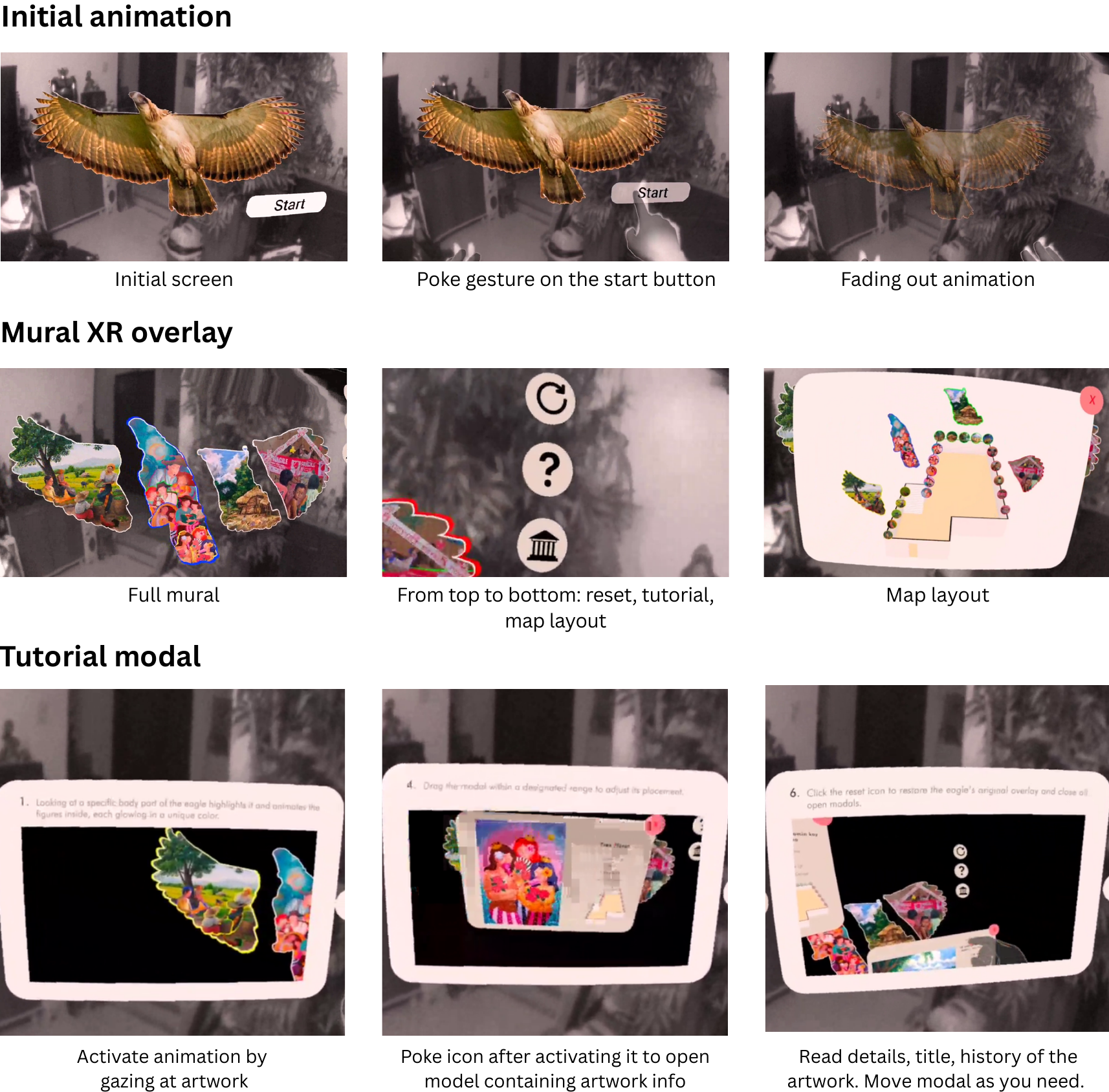}
\caption{Overview of the \textit{SoulWall} experience. Through gaze and touch-based interaction (\textit{galaw}), users activate artwork animations and information modals, supporting exploration and appreciation (\textit{gunita}) of the artworks. All featured artworks are used with the artists’ consent and remain their copyright.}
  \label{fig:teaser}
\end{teaserfigure}

\maketitle

\section{Introduction}
\label{sec:intro}

\par Digital and interactive systems are becoming increasingly common in cultural and public spaces, shaping how people encounter and engage with information. While these environments can be engaging, highly stimulating settings may also present challenges when content is dense, visually rich, or unstructured \cite{BaDeLang2012, barrouillet_law_2011, zeglen_increasing_2018, barrouillet_further_2011}. When multiple visual and sensory elements compete for attention, users may experience difficulty focusing on specific details or forming meaningful connections \cite{sweller2011cognitive}. Such conditions are common in public exhibition spaces, including museums and galleries, where visitors encounter a combination of artworks, text, spatial layouts, and ambient stimuli \shortcite{towse_recall_2008, varao-sousa_lab_2018}.

\par Museums and galleries are often designed to be rich and immersive, combining visual artifacts with explanatory text, digital media, and spatial narratives. While these elements aim to support interpretation and appreciation, they can also contribute to cognitive fatigue when presented simultaneously \cite{moreno_visual_1999, kotler_can_2007}. Prior work has shown that visitors may struggle to sustain attention or meaningfully engage with content in such environments, particularly when exhibits are dense or unfamiliar \cite{vuillemier_2005, phleps_2004, um2012}. These challenges raise questions about how interactive systems can support art appreciation without overwhelming visitors or detracting from the artworks themselves.

\par In parallel, digital technologies have opened new opportunities for presenting and experiencing art beyond traditional formats. Techniques such as visual layering, spatial organization, and interactive exploration have been used to support engagement and understanding across educational and cultural contexts \cite{mckenzie_improving_2003, qureshi_method_2014, mccabe_location_2015}. Interactive and immersive systems, including extended reality (XR), allow content to be arranged spatially and explored through embodied interaction, offering alternative ways of engaging with visual material \cite{familoni2024, cho2018spatial}. Large interactive wall displays, in particular, provide shared visual reference points that encourage exploration while maintaining a sense of scale and context \cite{jansen_effects_2019, netmarble_2020}.

\par In the Philippine context, visual art is closely tied to precolonial, colonial, and postcolonial histories, with many works carrying cultural, political, and social significance. At the same time, access to artworks is often limited by physical preservation concerns, exhibition constraints, or geographic availability. Rather than replacing physical and traditional art forms, digital systems can serve as complementary platforms that create interactive digital twins of artworks, allowing audiences to explore and engage with them without risking damage to the originals. Such approaches offer opportunities to support Filipino artists and broaden public access to their work while respecting artistic intent and material integrity.

\par In this paper, we present \textit{SoulWall}, an extended reality interactive mural system designed to augment Filipino artworks through digital animation and embodied interaction (see Figure~\ref{fig:teaser}). SoulWall, through the verbs \textit{Galaw} at \textit{Gunita}  (to touch and reminisce), allows viewers to explore paintings and animated elements, supporting playful and exploratory engagement while preserving the physical artworks. We describe the design and implementation of the system (see Figure~\ref{fig:soulwallall}), including its mural layout (see Figure~\ref{fig:mural}), interaction techniques, and interaction logging infrastructure. We also report findings from an initial evaluation focused on user experience, and supported by analyses of interaction logs and spatial engagement patterns. Together, these contributions position interactive XR murals as a viable approach for supporting art appreciation and showcasing Filipino artists in digital and public exhibition spaces. 

\section{Related Work}
\label{sec:related}

\par Prior work has explored how extended reality (XR) systems can support user engagement and orientation through immersive visualization, interactivity, and spatial organization \cite{zhang_exploring_2019, essoe_enhancing_2022}. AR and VR systems, particularly those using head-mounted displays, have been shown to support active exploration and sustained attention when compared to traditional presentation formats \cite{gargrish_measuring_2021, gargrish_evaluation_2022, zhang_virtuality_2023, krokos_virtual_2019, caluya2018transferability}. Gesture-based and embodied interactions further contribute to spatial awareness and intuitive navigation by allowing users to interact directly with digital content \cite{zagermann_memory_2017, lee_immersive_2018}. These approaches highlight the value of spatially situated and interactive representations for engaging users in information-rich environments.

\par Spatial structure and contextual anchoring also play an important role in how users navigate and make sense of digital content. Prior studies show that VR environments aligned with real-world layouts can support orientation more effectively than abstract representations \cite{miller_long-term_1999}. Similarly, mimetic icons, annotations, and spatial cues help users relate digital elements to physical or familiar reference points \cite{griffin_does_2005, fujimoto_relation_2012, kovacevic2022retzzles, kovacevic2023retzzles}. Techniques inspired by the method of loci emphasize the use of landmarks and spatial segmentation to organize content and guide exploration \cite{rosello_nevermind_2016, liu_investigating_2024}. While often discussed in the context of memory, these techniques also inform the design of navigable and interpretable interactive spaces. \textbf{Our work builds on these XR interaction and spatial organization principles by presenting artworks as part of an interactive mural}. Rather than using spatial structure to optimize recall, our system applies these techniques to support exploration, orientation, and engagement with digital representations of Filipino artworks. By anchoring animations and interactions directly to visual elements within the mural, the system encourages viewers to move, observe, and interact with the artworks in a manner that remains grounded in their visual and spatial context.

\par Museums and galleries present unique challenges and opportunities for interactive systems, as environmental factors such as layout, visibility, and distractions influence how visitors engage with exhibits \cite{krukar2014walk, krukar2015influence}. Prior work suggests that prolonged engagement and active interaction with exhibits can support deeper appreciation and understanding \cite{sweetman2020material}. XR systems have been introduced in museum contexts to increase visitor engagement, usability, and perceived value, with immersion, enjoyment, and curiosity playing important roles in shaping visitor experience \cite{pallud_impact_2017}. Virtual environments that present related objects or narratives together have also been shown to encourage exploratory behavior and informal learning \cite{moesgaard_implicit_2015}. Large-scale deployments, such as interactive historical exhibits, further demonstrate the potential of digital systems to broaden access to cultural content \cite{mcgaughey_evaluating_2021}. \textbf{In contrast to systems that prioritize informational delivery or learning outcomes, our work positions XR as a complementary medium for art appreciation especially towards deeper appreciation for local Filipino art}. By creating an interactive digital twin of physical artworks, the system allows visitors to engage with Filipino art through movement, touch, and visual exploration without altering or risking damage to the originals. This work extends prior research on XR in museums by focusing on user experience and interaction patterns in an interactive mural setting, emphasizing accessibility, engagement, and respect for traditional art forms.

\section{Features of SoulWall}
\label{sec:features}


\begin{figure*}
    \centering
    \includegraphics[width=\linewidth]{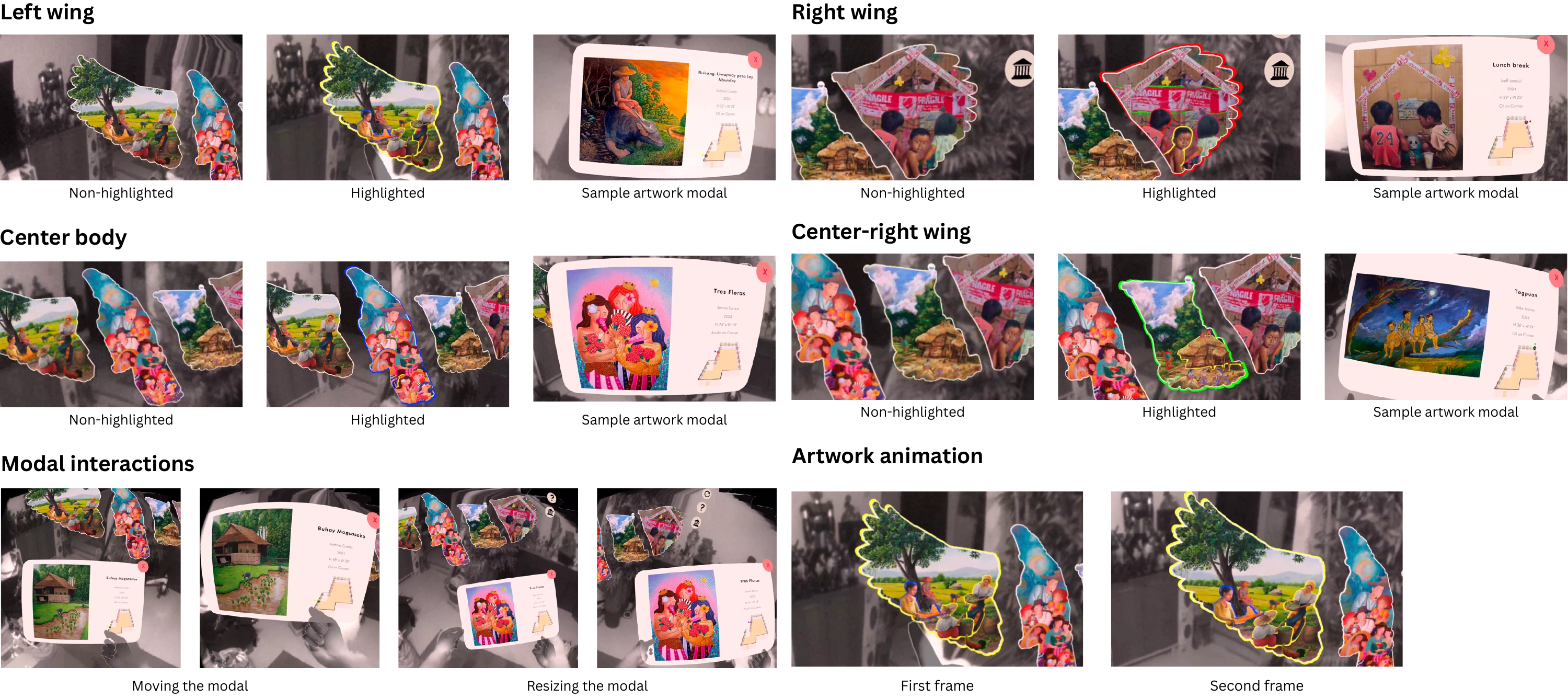}
    \caption{Mural segment features of SoulWall, showing interactive artwork cues, highlighted and non-highlighted states across segments, and available modal interactions. Top Row: We show the murals from both the Left and Right side wing of the mural. Middle Row: We show the murals from the Center body and Center-right wing of the mural. Bottom RoW: We show the different interactions applicable with both the modals and the animations behind some of the paintings.}
    \label{fig:soulwallall}
\end{figure*}

\par SoulWall integrates a set of interactive features designed to support the exploration and appreciation of Filipino artworks through an extended reality (XR) mural. Rather than replacing physical artworks, the system augments them by creating an interactive digital twin that allows viewers to engage with visual elements at scale while preserving the integrity of the originals. The following features focus on spatial organization, visual cues, and lightweight interactions that encourage walk-up use and exploratory engagement.

\subsection{Mural Design}

\par The mural uses the Philippine eagle as an abstract and symbolic visual anchor that aligns with Filipino cultural identity. The eagle serves as a unifying visual structure that organizes the artworks within a single composition, helping viewers orient themselves and explore different sections of the mural in a coherent manner. Spatially structured XR environments have been shown to support navigation and engagement by providing clear visual reference points \cite{fujimoto_relation_2012, moll_optimized_2022, liu_investigating_2024}.

\par
\begin{figure}[t]
    \centering
    \includegraphics[width=1\linewidth]{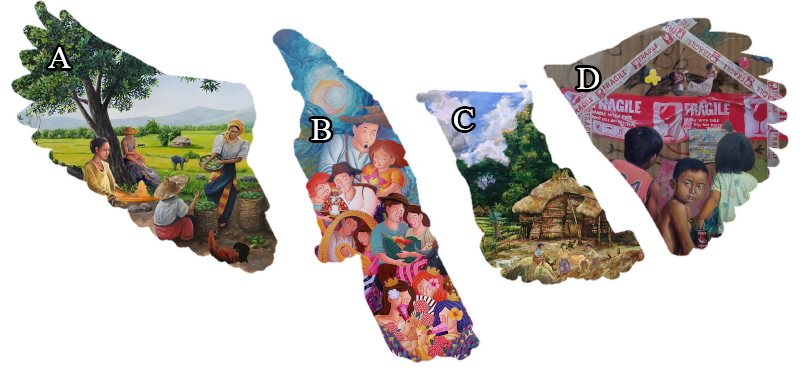}
    \caption{Eagle mural layout used as the visual structure for SoulWall. The paintings are grouped per body part which also depict a specific historical era of the country from the oldest (leftmost) to the most recent (rightmost).}
    \label{fig:mural}
\end{figure}

 The eagle is divided into four segments—A) left wing, B) center body, C) center-right wing, and D) right wing—physically printed on a large print (Figure~\ref{fig:mural}). This segmentation allows artworks to be grouped into visually distinct regions, enabling viewers to focus on specific parts of the mural without being overwhelmed by the entire display at once. Such spatial grouping supports gradual exploration and encourages viewers to move closer to different sections of the mural.

\par Each eagle segment contains a curated set of artworks from a featured Filipino artist. The artworks are arranged to maintain visual cohesion and thematic consistency within each segment, allowing viewers to appreciate stylistic elements and recurring motifs. By embedding the artworks within a single mural composition, SoulWall provides a shared visual context that supports comparison, reflection, and discussion while respecting the original works \cite{mckenzie_improving_2003}.

\subsection{Interactions}

\par SoulWall employs simple, trigger-based interactions that bring digital artworks to life through animation and visual feedback. These interactions are designed to enhance engagement and curiosity while keeping interaction demands low, allowing viewers to focus on the artworks themselves rather than on learning complex controls \cite{zhang_exploring_2019, xie_more_2017}. Subtle animations are used to indicate interactivity and to draw attention to specific visual elements without altering the core composition of the artworks \cite{jangid_contribution_2023, essoe_enhancing_2022}.

\par Prior work suggests that users prefer familiar and easily interpretable gestures when encountering XR systems for the first time \cite{gavgiotaki2023gesture}. Guided by this, SoulWall adopts interaction techniques that align with common real-world metaphors and interface conventions. The supported interactions include:

\begin{enumerate}
  \item \textbf{Gaze-Based Highlighting:} Looking at a specific segment of the eagle highlights it and triggers subtle animations on embedded figures, indicating that the artworks are interactive.
  \item \textbf{Modal Activation:} Selecting a highlighted figure opens a modal containing contextual information about the selected artwork and artist.
  \item \textbf{Modal Dismissal:} The modal can be closed using a standard close button.
  \item \textbf{Modal Repositioning:} Users may drag the modal within a bounded area to adjust its placement for comfortable viewing.
  \item \textbf{Pinch-to-Zoom:} Pinch gestures allow users to zoom in on details or zoom out for an overview of the mural or modal content.
  \item \textbf{Reset Function:} A reset icon restores the mural to its default state and closes all open modals.
  \item \textbf{Help Access:} A help icon replays a short tutorial that explains the available interactions.
  \item \textbf{Scene Switching:} A museum icon transitions the view from the mural to an overview of the exhibition layout.
\end{enumerate}

\par Together, these interactions encourage viewers to move, observe, and engage with the artworks at their own pace. By combining spatial organization with lightweight interaction, SoulWall supports exploratory art appreciation while minimizing disruption to the visual integrity of the mural. An overview of the interaction flow is shown in Figure~\ref{fig:teaser}, and examples of mural segments and artwork modals are shown in Figure~\ref{fig:soulwallall}.



\section{Evaluation}
\label{sec:evaluation}

\subsection{Participants and Protocol}

\par Sixteen participants (aged 18–25) were recruited through convenience sampling. All participants provided informed consent prior to participation. The study focused on evaluating user experience and interaction behavior during an exploratory session with the SoulWall prototype.

\par Participants were first given a brief introduction to the system and its available interactions. They were then asked to freely explore the interactive mural for approximately 15 minutes. No specific tasks or performance goals were imposed in order to encourage natural and self-directed interaction with the artworks. After the exploration phase, participants completed the short version of the User Experience Questionnaire (UEQ-S). During the session, interaction logs were collected automatically by the system, capturing movement, interaction events, and dwell patterns (similar to that of approach of ~\cite{kovacevic2022retzzles}). Basic background information related to prior XR experience and handedness was also recorded for descriptive purposes.

\subsection{Experimental Setup}

\par The study was conducted in an indoor exhibition-like space where the mural was physically installed as a large printed artwork. Participants interacted with the augmented mural using a standalone XR headset. The setup allowed participants to stand in front of the mural and move freely while exploring the interactive elements.

\par A facilitator was present to assist with basic setup and to address technical issues if needed, but did not guide participant behavior during exploration. The participant’s view was mirrored to a laptop for observation and to ensure correct system operation.

\subsection{Findings}
\begin{figure}
    \centering
    \includegraphics[width=\columnwidth]{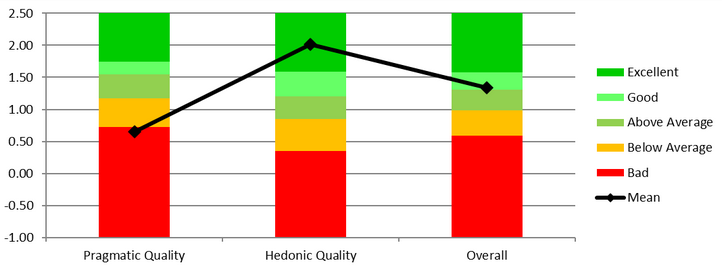}
    \caption{UEQ-S results for SoulWall, showing mean scores for pragmatic quality, hedonic quality, and overall user experience. Colored bands indicate benchmark categories from bad to excellent, with the black line representing mean values. Graph was scaled to half values since we used the short version of the UEQ. Score trasmutations were made with the official analysis tool by ~\cite{schrepp2017design}).}
    \label{fig:ueq}
\end{figure}

\begin{figure}
    \centering
    \includegraphics[width=\columnwidth]{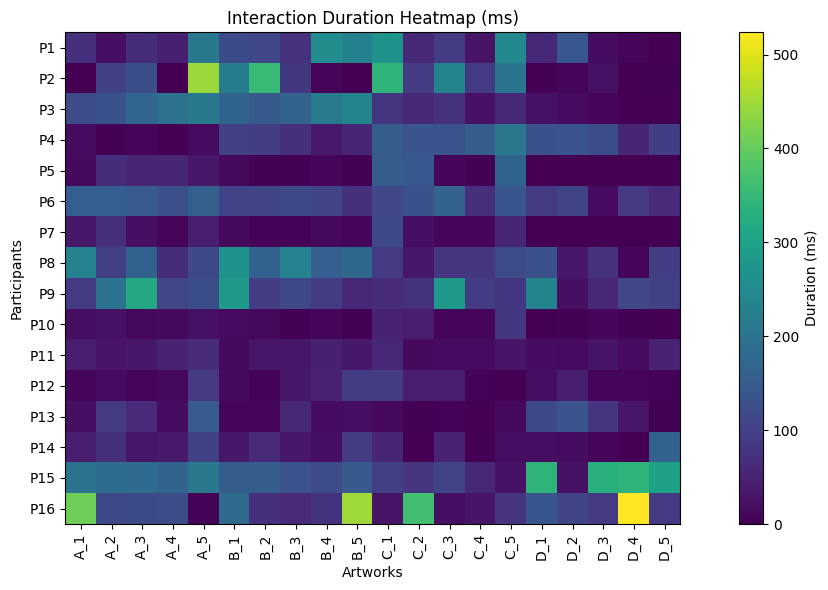}
    \caption{Heatmap of interaction duration (in milliseconds) across participants and artworks. Rows represent participants and columns represent individual artworks. Higher intensity indicates longer engagement time with an artwork.}
    \label{fig:engagement}
\end{figure}

\par User experience (as seen in ~\autoref{fig:ueq}) was assessed using the UEQ-S, which measures pragmatic and hedonic qualities. For SoulWall, pragmatic quality received a mean score of $M = 4.66$ ($SD = 0.61$), indicating that participants generally found the system understandable and manageable during short-term use. Hedonic quality was rated higher, with a mean score of $M = 6.02$ ($SD = 0.71$), suggesting that the interactive mural was perceived as engaging and enjoyable. The overall UEQ-S score for SoulWall was $M = 5.34$ ($SD = 0.54$), reflecting a generally positive user experience that supported exploratory interaction without requiring extensive instruction.

\par Interaction logs provided additional insights into how participants engaged with the mural (see ~\autoref{fig:engagement}). Logged data included movement trajectories, time spent near different mural segments, and interaction events such as selections and modal activations. Across artworks, movement activity was highest for artwork C\_5 (M = 15.63 movements, SD = 35.14), followed by B\_1 (M = 15.31, SD = 47.82) and A\_3 (M = 15.19, SD = 39.36). Visualizations of these logs showed that participants explored multiple sections of the mural rather than remaining in a single area, with noticeable clustering around visually dense or animated regions. This suggests that animations and visual cues effectively drew attention and encouraged movement across the mural.

\par Interaction duration revealed a different pattern of engagement. Artwork A\_5 received the longest interaction times on average (M = 229.57 ms, SD = 454.74 ms), followed by C\_1 (M = 203.90 ms, SD = 401.54 ms) and B\_1 (M = 195.50 ms, SD = 387.81 ms). At the participant level, interaction patterns varied substantially. Participant P16 exhibited the highest overall movement activity across artworks, while participant P15 accumulated the greatest total interaction duration. These differences reflect distinct exploration styles, with some participants engaging through frequent movement across artworks and others spending longer periods inspecting individual pieces and associated information modals. Together, these findings indicate that SoulWall supported both focused and broad exploration strategies, allowing users to engage with the artworks at their own pace.

\par Taken together, the UEQ-S results and interaction logs indicate that SoulWall supported a positive and flexible user experience for engaging with Filipino artworks through an interactive XR mural. Hedonic quality scores reflected high enjoyment, while interaction logs showed varied engagement patterns across artworks and participants, with some artworks receiving higher movement activity and others longer interaction durations. Rather than optimizing for task performance, SoulWall facilitated open-ended exploration and visual engagement, allowing participants to interact with artworks in different ways. These findings align with the system’s goal of complementing traditional art displays by enabling interactive appreciation without altering or replacing the original works.

\section{Discussion and Design Implications}
\label{sec:discussion}

\par We synthesize the findings from the evaluation with existing work on XR interaction techniques, spatial organization, and museum-based interactive systems, and discuss how these insights can inform the design of XR murals that support exploratory and respectful engagement with artworks \cite{zhang_exploring_2019, jansen_effects_2019, pallud_impact_2017}.

\subsection{Designing for visual salience in interactive murals}

\par Interaction logs indicate that visual salience played a central role in shaping how participants engaged with the SoulWall mural. Artworks that were visually larger, more centrally positioned, or featured animated elements tended to attract higher interaction activity, either through longer engagement durations or more frequent movement events. In contrast, smaller or visually dense artworks were interacted with less consistently across participants. These patterns align with prior findings showing that visual prominence strongly guides attention and exploration in complex visual environments \shortcite{rayner1992, henderson1999}.

\par Importantly, the interaction logs revealed that high movement activity and long interaction duration did not always occur on the same artworks. Some artworks elicited frequent movement across participants, while others encouraged longer periods of focused engagement. This distinction suggests that visual salience can support different modes of engagement prompting either exploratory movement or sustained inspection rather than a single dominant interaction pattern. These observations are consistent with XR and display-based studies showing that perceptual features often influence how users engage, independent of the interaction techniques available \cite{zhang_exploring_2019, jansen_effects_2019}.

\par For XR mural design, these findings highlight the need to balance artistic composition with interaction visibility. Subtle animations and visual cues can draw attention and invite exploration without overwhelming the artwork, while overly subtle or crowded elements may limit discoverability. Designing for visual salience can therefore support varied forms of engagement while preserving the integrity of the artworks.

\subsection{Supporting spatial exploration through mural structure}

\par Prior work on XR navigation and interactive wall displays emphasizes the importance of spatial structure in supporting orientation and exploration \cite{miller_long-term_1999, jansen_effects_2019}. In SoulWall, the use of a single cohesive mural provided a shared visual reference that encouraged participants to traverse multiple mural regions rather than remain focused on a single area. Interaction logs showed movement across different segments, with localized clustering around particular artworks before participants transitioned to adjacent areas.

\par At the same time, engagement patterns suggest that participants relied more heavily on visually salient cues than on the intended thematic segmentation of the mural. While the mural structure supported movement across space, boundaries between segments were not always explicitly reflected in interaction behavior. Similar challenges have been reported in XR environments where spatial metaphors are present but not sufficiently reinforced through visual or interaction cues \cite{tversky2003, griffin_does_2005}.

\par These observations suggest that interactive murals benefit from spatial structures that are both meaningful and perceptually legible. Reinforcing segmentation through visual framing, spacing, or background variation may help users better understand how different regions relate to one another, supporting smoother and more intentional spatial exploration.

\subsection{Balancing engagement and interaction simplicity}

\par Consistent with prior work on immersive museum experiences, participants reported high hedonic quality for SoulWall, indicating that animation, scale, and embodied interaction contributed to an engaging experience \cite{pallud_impact_2017, sweetman2020material}. UEQ-S results suggest that the system successfully supported enjoyable and playful interaction, which is important for public-facing cultural installations.

\par Interaction logs further revealed substantial variation in how participants engaged with the mural. Some participants accumulated high movement counts across many artworks, while others spent longer durations engaging with fewer pieces. This variation reflects different exploration strategies, ranging from broad scanning of the mural to more focused inspection of individual artworks and information modals. Such diversity aligns with prior findings that users adopt distinct engagement styles in open-ended XR environments \cite{cho2018spatial}.

\par These findings reinforce the importance of keeping interactions lightweight and optional. Prior work has cautioned that overly complex interactions can distract from content in immersive systems \cite{makransky2019}. In SoulWall, familiar gestures and minimal interaction steps appeared to support engagement without enforcing a single interaction style, allowing participants to navigate the experience according to their preferences.

\subsection{Familiarity and onboarding in walk-up XR experiences}

\par Familiar interaction metaphors have been shown to support usability in XR systems \cite{gavgiotaki2023gesture, kim_investigation_2020}. In SoulWall, participants were generally able to understand gaze-based highlighting, selection, and modal interactions with limited instruction, suggesting that familiarity helped lower the initial barrier to engagement.

\par However, interaction logs and observations indicate that some features were underutilized when participants skipped or quickly dismissed the tutorial. This reflects a common challenge in walk-up-and-use XR installations, where users are eager to explore but may avoid explicit onboarding \cite{jansen_effects_2019}. Integrating guidance directly into the interaction flow through progressive cues or contextual highlights, may better support first-time users without interrupting exploration.

\par Overall, these findings echo prior work on interactive museum systems, which emphasizes keeping technology unobtrusive and allowing visitors to engage at their own pace \cite{pallud_impact_2017, moesgaard_implicit_2015}. For XR murals, familiarity in interaction design and lightweight onboarding help ensure that attention remains on the artworks rather than on the interface.

\par Taken together, the discussion reinforces that effective XR murals for art appreciation should prioritize visual clarity, spatial legibility, and interaction simplicity. By grounding design decisions in both prior research and observed interaction patterns, such as differences between movement-driven and duration-driven engagement, systems like SoulWall can complement traditional art displays while supporting diverse and self-directed forms of interaction with cultural content.

\section{Future Work and Conclusion}
\label{sec:conclusion}

\par This paper presented SoulWall, an XR interactive mural designed to support the exploration and appreciation of Filipino artworks through digital augmentation. Rather than replacing physical artworks, SoulWall creates an interactive digital twin that allows viewers to engage (\textit{galaw}) with visual elements while preserving the integrity of the originals and be able to appreciate (\textit{gunita}) the art they are interacting with. We described the system design and reported findings from an evaluation focused on user experience and interaction behavior.

\par Results from the UEQ-S and interaction logs suggest that SoulWall provides an engaging and flexible experience, supporting different exploration styles through lightweight interactions and spatial organization. Participants were drawn to visually salient and animated elements, and interaction patterns indicated movement across multiple mural regions rather than fixed or linear use. These findings highlight the importance of visual clarity, spatial legibility, and simple interaction design in XR-based art installations.

\par This work has several limitations. The study focused on short-term use with a relatively homogeneous participant group and was conducted in a controlled setting rather than a live exhibition environment. As a result, the findings primarily reflect initial engagement and exploratory behavior.

\par Future work will examine SoulWall in longer-term and in-situ deployments, such as museums or public cultural spaces, and with more diverse audiences. We also plan to explore how interactive XR murals can support deeper forms of reflection and understanding, including spatial memory and recall, which are beyond the scope of the present paper. Overall, this work contributes an initial step toward designing XR murals that complement traditional art displays and expand access to Filipino art through interactive digital experiences.

\begin{acks}
\par We would like to thank the participating Filipino artists (Joeff Jamalul, Janina Sanico, Antonio Cuerdo, Mike Jacinto) for generously allowing us to use their artworks in this project. Their consent made it possible to explore new ways of promoting and experiencing Filipino art through interactive digital systems. We are grateful for their trust and support in this project. 
\end{acks}

\bibliographystyle{ACM-Reference-Format}
\bibliography{references}

@article{kim_investigation_2020,
	title = {Investigation of {Delayed} {Response} during {Real}-{Time} {Cursor} {Control} {Using} {Electroencephalography}},
	volume = {2020},
	copyright = {Copyright {\textcopyright} 2020 Hyeonseok Kim et al.},
	issn = {2040-2309},
	url = {https://onlinelibrary.wiley.com/doi/abs/10.1155/2020/1418437},
	doi = {10.1155/2020/1418437},
	language = {en},
	number = {1},
	urldate = {2025-07-17},
	journal = {Journal of Healthcare Engineering},
	author = {Kim, Hyeonseok and Yoshimura, Natsue and Koike, Yasuharu},
	year = {2020},
	note = {\_eprint: https://onlinelibrary.wiley.com/doi/pdf/10.1155/2020/1418437},
	pages = {1418437},
}

@article{essoe_enhancing_2022,
	title = {Enhancing learning and retention with distinctive virtual reality environments and mental context reinstatement},
	volume = {7},
	copyright = {2022 The Author(s)},
	issn = {2056-7936},
	url = {https://www.nature.com/articles/s41539-022-00147-6},
	doi = {10.1038/s41539-022-00147-6},
	language = {en},
	number = {1},
	urldate = {2024-06-26},
	journal = {npj Science of Learning},
	author = {Essoe, Joey Ka-Yee and Reggente, Nicco and Ohno, Ai Aileen and Baek, Younji Hera and Dell’Italia, John and Rissman, Jesse},
	month = dec,
	year = {2022},
	note = {Publisher: Nature Publishing Group},
	pages = {1--14},
}

@article{gargrish_evaluation_2022,
	title = {Evaluation of memory retention among students using augmented reality based geometry learning assistant},
	volume = {27},
	issn = {1573-7608},
	url = {https://doi.org/10.1007/s10639-022-11147-9},
	doi = {10.1007/s10639-022-11147-9},
	language = {en},
	number = {9},
	urldate = {2024-06-26},
	journal = {Education and Information Technologies},
	author = {Gargrish, Shubham and Mantri, Archana and Kaur, Deepti Prit},
	month = nov,
	year = {2022},
	pages = {12891--12912},
}

@incollection{kotler_can_2007,
	title = {Can {Museums} be {All} {Things} to {All} {People}?: {Missions}, goals, and marketing's role},
	isbn = {978-0-203-96419-4},
	shorttitle = {Can {Museums} be {All} {Things} to {All} {People}?},
	booktitle = {Museum {Management} and {Marketing}},
	publisher = {Routledge},
	author = {Kotler, Neil and Kotler, Philip},
	year = {2007},
	note = {Num Pages: 18},
}

@inproceedings{zhang_exploring_2019,
	title = {Exploring {Effects} of {Interactivity} on {Learning} with {Interactive} {Storytelling} in {Immersive} {Virtual} {Reality}},
	url = {https://ieeexplore.ieee.org/document/8864531},
	doi = {10.1109/VS-Games.2019.8864531},
	urldate = {2024-06-26},
	booktitle = {2019 11th {International} {Conference} on {Virtual} {Worlds} and {Games} for {Serious} {Applications} ({VS}-{Games})},
	author = {Zhang, Lei and Bowman, Doug A. and Jones, Caroline N.},
	month = sep,
	year = {2019},
	note = {ISSN: 2474-0489},
	pages = {1--8},
}

@article{griffin_does_2005,
	title = {Does spatial or visual information in maps facilitate text recall? {Reconsidering} the conjoint retention hypothesis},
	volume = {53},
	issn = {1556-6501},
	shorttitle = {Does spatial or visual information in maps facilitate text recall?},
	url = {https://doi.org/10.1007/BF02504855},
	doi = {10.1007/BF02504855},
	language = {en},
	number = {1},
	urldate = {2024-06-26},
	journal = {Educational Technology Research and Development},
	author = {Griffin, Marlynn M. and Robinson, Daniel H.},
	month = mar,
	year = {2005},
	pages = {23--36},
}

@inproceedings{zhang_virtuality_2023,
	address = {New York, NY, USA},
	series = {{EICS} '23 {Companion}},
	title = {Virtuality or {Physicality}? {Supporting} {Memorization} {Through} {Augmented} {Reality} {Gamification}},
	isbn = {9798400702068},
	shorttitle = {Virtuality or {Physicality}?},
	url = {https://doi.org/10.1145/3596454.3597183},
	doi = {10.1145/3596454.3597183},
	urldate = {2024-06-26},
	booktitle = {Companion {Proceedings} of the 2023 {ACM} {SIGCHI} {Symposium} on {Engineering} {Interactive} {Computing} {Systems}},
	publisher = {Association for Computing Machinery},
	author = {Zhang, Yuchong and Nowak, Adam and Romanowski, Andrzej and Fjeld, Morten},
	month = jun,
	year = {2023},
	pages = {53--58},
}

@inproceedings{zagermann_memory_2017,
	address = {New York, NY, USA},
	series = {{CHI} '17},
	title = {Memory in {Motion}: {The} {Influence} of {Gesture}- and {Touch}-{Based} {Input} {Modalities} on {Spatial} {Memory}},
	isbn = {978-1-4503-4655-9},
	shorttitle = {Memory in {Motion}},
	url = {https://doi.org/10.1145/3025453.3026001},
	doi = {10.1145/3025453.3026001},
	urldate = {2024-06-26},
	booktitle = {Proceedings of the 2017 {CHI} {Conference} on {Human} {Factors} in {Computing} {Systems}},
	publisher = {Association for Computing Machinery},
	author = {Zagermann, Johannes and Pfeil, Ulrike and Fink, Daniel and von Bauer, Philipp and Reiterer, Harald},
	month = may,
	year = {2017},
	pages = {1899--1910},
}

@article{kovacevic2023retzzles,
  author    = {Kovacevic, N. and Deja, J. A. and Weerasinghe, M. and Copic Pucihar, K. and Kljun, M.},
  title     = {Retzzles: Towards Supporting Retention Using Puzzle Interactions and Abstract Symbols},
  journal = {Scientific Presentations of Primatijada 2023},
  year      = {2023},
  month     = {April 28--May 03},
  address   = {Ohrid, North Macedonia},
  pages     = {1--5}
}

@inproceedings{fujimoto_relation_2012,
	address = {New York, NY, USA},
	series = {{AH} '12},
	title = {Relation between location of information displayed by augmented reality and user's memorization},
	isbn = {978-1-4503-1077-2},
	url = {https://doi.org/10.1145/2160125.2160132},
	doi = {10.1145/2160125.2160132},
	urldate = {2024-06-26},
	booktitle = {Proceedings of the 3rd {Augmented} {Human} {International} {Conference}},
	publisher = {Association for Computing Machinery},
	author = {Fujimoto, Yuichiro and Yamamoto, Goshiro and Kato, Hirokazu and Miyazaki, Jun},
	month = mar,
	year = {2012},
	pages = {1--8},
}

@article{miller_long-term_1999,
	title = {Long-{Term} {Retention} of {Spatial} {Knowledge} {Acquired} in {Virtual} {Reality}},
	volume = {43},
	issn = {1071-1813},
	url = {https://doi.org/10.1177/154193129904302216},
	doi = {10.1177/154193129904302216},
	language = {en},
	number = {22},
	urldate = {2024-06-26},
	journal = {Proceedings of the Human Factors and Ergonomics Society Annual Meeting},
	author = {Miller, Michael S. and Clawson, Deborah M. and Sebrechts, Marc M.},
	month = sep,
	year = {1999},
	note = {Publisher: SAGE Publications Inc},
	pages = {1243--1246},
}

@article{gargrish_measuring_2021,
	title = {Measuring effectiveness of augmented reality-based geometry learning assistant on memory retention abilities of the students in {3D} geometry},
	volume = {29},
	copyright = {© 2021 Wiley Periodicals LLC},
	issn = {1099-0542},
	url = {https://onlinelibrary.wiley.com/doi/abs/10.1002/cae.22424},
	doi = {10.1002/cae.22424},
	language = {en},
	number = {6},
	urldate = {2024-06-26},
	journal = {Computer Applications in Engineering Education},
	author = {Gargrish, Shubham and Kaur, Deepti P. and Mantri, Archana and Singh, Gurjinder and Sharma, Bhanu},
	year = {2021},
	note = {\_eprint: https://onlinelibrary.wiley.com/doi/pdf/10.1002/cae.22424},
	pages = {1811--1824},
}

@article{lee_immersive_2018,
	title = {Immersive Gesture Interfaces for Navigation of 3D Maps in {HMD}-Based Mobile Virtual Environments},
	volume = {2018},
	rights = {Copyright © 2018 Yea Som Lee and Bong-Soo Sohn.},
	issn = {1875-905X},
	url = {https://onlinelibrary.wiley.com/doi/abs/10.1155/2018/2585797},
	doi = {10.1155/2018/2585797},
	pages = {2585797},
	number = {1},
	journal = {Mobile Information Systems},
	author = {Lee, Yea Som and Sohn, Bong-Soo},
	urldate = {2024-06-07},
	year = {2018},
	langid = {english},
	note = {\_eprint: https://onlinelibrary.wiley.com/doi/pdf/10.1155/2018/2585797},
}

@article{pallud_impact_2017,
	title = {Impact of interactive technologies on stimulating learning experiences in a museum},
	volume = {54},
	issn = {0378-7206},
	url = {https://www.sciencedirect.com/science/article/pii/S0378720616302634},
	doi = {10.1016/j.im.2016.10.004},
	number = {4},
	urldate = {2024-10-02},
	journal = {Information \& Management},
	author = {Pallud, Jessie},
	month = jun,
	year = {2017},
	pages = {465--478},
}

@article{moesgaard_implicit_2015,
	title = {Implicit and {Explicit} {Information} {Mediation} in a {Virtual} {Reality} {Museum} {Installation} and its {Effects} on {Retention} and {Learning} {Outcomes}: {The} 9th {European} {Conference} on {Games} {Based} {Learning} {ECGBL} 2015},
	issn = {978-1-910810-58-3��},
	shorttitle = {Implicit and {Explicit} {Information} {Mediation} in a {Virtual} {Reality} {Museum} {Installation} and its {Effects} on {Retention} and {Learning} {Outcomes}},
	journal = {Proceedings of The 9th European Conference on Games-Based Learning},
	author = {Moesgaard, Tomas Gislason and Witt, Mass and Fiss, Jonathan and Warming, Cecilie and Klubien, Jonathan and Schønau-Fog, Henrik},
	editor = {Munkvold, Robin and Kolås, Line},
	month = oct,
	year = {2015},
	note = {Place: Reading, United Kingdom
Publisher: Academic Conferences and Publishing International},
	pages = {387--394},
}

@article{towse_recall_2008,
	title = {The recall of information from working memory: insights from behavioural and chronometric perspectives},
	volume = {55},
	issn = {1618-3169},
	shorttitle = {The recall of information from working memory},
	url = {https://www.ncbi.nlm.nih.gov/pmc/articles/PMC2658622/},
	number = {6},
	urldate = {2024-10-12},
	journal = {Experimental psychology},
	author = {Towse, John N. and Cowan, Nelson and Hitch, Graham J. and Horton, Neil J},
	year = {2008},
	pmid = {19130763},
	pmcid = {PMC2658622},
	pages = {371--383},
}

@article{varao-sousa_lab_2018,
	title = {In the lab and in the wild: {How} distraction and mind wandering affect attention and memory},
	volume = {3},
	issn = {2365-7464},
	shorttitle = {In the lab and in the wild},
	url = {https://doi.org/10.1186/s41235-018-0137-0},
	doi = {10.1186/s41235-018-0137-0},
	number = {1},
	urldate = {2024-10-12},
	journal = {Cognitive Research: Principles and Implications},
	author = {Varao-Sousa, Trish L. and Smilek, Daniel and Kingstone, Alan},
	month = nov,
	year = {2018},
	pages = {42},
}

@misc{mcgaughey_evaluating_2021,
	title = {Evaluating the {Impact} of {Augmented} and {Virtual} {Reality}},
	url = {https://www.aam-us.org/2021/01/15/evaluating-the-impact-of-augmented-and-virtual-reality/},
	language = {en-US},
	urldate = {2024-10-02},
	journal = {American Alliance of Museums},
	author = {McGaughey, Colleen and Russick, John},
	month = jan,
	year = {2021},
}

@article{krokos_virtual_2019,
	title = {Virtual memory palaces: immersion aids recall},
	volume = {23},
	issn = {1434-9957},
	shorttitle = {Virtual memory palaces},
	url = {https://doi.org/10.1007/s10055-018-0346-3},
	doi = {10.1007/s10055-018-0346-3},
	language = {en},
	number = {1},
	urldate = {2024-09-12},
	journal = {Virtual Reality},
	author = {Krokos, Eric and Plaisant, Catherine and Varshney, Amitabh},
	month = mar,
	year = {2019},
	pages = {1--15},
}

@article {BaDeLang2012,
author = {Barrouillet, P. and De Paepe, A. and Langerock, N.},
title = "Time causes forgetting from working memory",
journal = "Psychonomic Bulletin \& Review",
volume = "19",
number = "1",
pages = "87-92",
year = "2012"
}

@article{barrouillet_law_2011,
	title = {On the law relating processing to storage in working memory.},
	volume = {118},
	issn = {1939-1471, 0033-295X},
	url = {https://doi.apa.org/doi/10.1037/a0022324},
	doi = {10.1037/a0022324},
	language = {en},
	number = {2},
	urldate = {2024-11-09},
	journal = {Psychological Review},
	author = {Barrouillet, Pierre and Portrat, Sophie and Camos, Valérie},
	year = {2011},
	pages = {175--192},
}

@book{liu_investigating_2024,
	title = {Investigating the {Effects} of {Physical} {Landmarks} on {Spatial} {Memory} for {Information} {Visualisation} in {Augmented} {Reality}},
	author = {Liu, Jiazhou and Satriadi, Kadek and Ens, Barrett and Dwyer, Tim},
	month = sep,
	year = {2024},
	doi = {10.13140/RG.2.2.29017.94563},
}

@inproceedings{rosello_nevermind_2016,
	address = {Tokyo Japan},
	title = {{NeverMind}: {Using} {Augmented} {Reality} for {Memorization}},
	copyright = {http://www.acm.org/publications/policies/copyright\_policy\#Background},
	isbn = {978-1-4503-4531-6},
	shorttitle = {{NeverMind}},
	url = {https://dl.acm.org/doi/10.1145/2984751.2984776},
	doi = {10.1145/2984751.2984776},
	language = {en},
	urldate = {2024-11-09},
	booktitle = {Proceedings of the 29th {Annual} {Symposium} on {User} {Interface} {Software} and {Technology}},
	publisher = {ACM},
	author = {Rosello, Oscar and Exposito, Marc and Maes, Pattie},
	month = oct,
	year = {2016},
	pages = {215--216},
}

@article{mccabe_location_2015,
	title = {Location, {Location}, {Location}! {Demonstrating} the {Mnemonic} {Benefit} of the {Method} of {Loci}},
	volume = {42},
	issn = {0098-6283},
	url = {https://doi.org/10.1177/0098628315573143},
	doi = {10.1177/0098628315573143},
	language = {en},
	number = {2},
	urldate = {2024-11-20},
	journal = {Teaching of Psychology},
	author = {McCabe, Jennifer A.},
	month = apr,
	year = {2015},
	note = {Publisher: SAGE Publications Inc},
	pages = {169--173},
}

@article{mckenzie_improving_2003,
	title = {Improving {Comprehension} through {Mural} {Lessons}},
	volume = {56},
	issn = {0034-0561},
	url = {https://www.jstor.org/stable/20205287},
	number = {8},
	urldate = {2024-11-20},
	journal = {The Reading Teacher},
	author = {McKenzie, Gary R. and Danielson, Elaine},
	year = {2003},
	note = {Publisher: [Wiley, International Reading Association]},
	pages = {738--742},
}

@inproceedings{jansen_effects_2019,
	address = {New York, NY, USA},
	series = {{CHI} '19},
	title = {Effects of {Locomotion} and {Visual} {Overview} on {Spatial} {Memory} when {Interacting} with {Wall} {Displays}},
	isbn = {978-1-4503-5970-2},
	url = {https://doi.org/10.1145/3290605.3300521},
	doi = {10.1145/3290605.3300521},
	urldate = {2024-11-20},
	booktitle = {Proceedings of the 2019 {CHI} {Conference} on {Human} {Factors} in {Computing} {Systems}},
	publisher = {Association for Computing Machinery},
	author = {Jansen, Yvonne and Schjerlund, Jonas and Hornbæk, Kasper},
	month = may,
	year = {2019},
	pages = {1--12},
}

@misc{netmarble_2020,
	title = {NetMarble Game Academy Exhibition},
	url = {https://www.img-lab.co.kr/10},
	language = {ko},
        year = {2020},
author={IMG-LAB INC.},
	urldate = {2024-11-20},
	journal = {img-lab},
}

@article{qureshi_method_2014,
	title = {The method of loci as a mnemonic device to facilitate learning in endocrinology leads to improvement in student performance as measured by assessments},
	volume = {38},
	url = {https://pmc.ncbi.nlm.nih.gov/articles/PMC4056179/},
	doi = {10.1152/advan.00092.2013},
	language = {en},
	number = {2},
	urldate = {2024-11-09},
	journal = {Advances in Physiology Education},
	author = {Qureshi, Ayisha and Rizvi, Farwa and Syed, Anjum and Shahid, Aqueel and Manzoor, Hana},
	month = jun,
	year = {2014},
	pmid = {25039085},
	pages = {140},
}

@article{barrouillet_further_2011,
	title = {Further {Evidence} for {Temporal} {Decay} in {Working} {Memory}: {Reply} to {Lewandowsky} and {Oberauer} (2009)},
	volume = {37},
	shorttitle = {Further {Evidence} for {Temporal} {Decay} in {Working} {Memory}},
	doi = {10.1037/a0022933},
	journal = {Journal of experimental psychology. Learning, memory, and cognition},
	author = {Barrouillet, Pierre and Portrat, Sophie and Vergauwe, Evie and Diependaele, Kevin and Camos, Valérie},
	month = sep,
	year = {2011},
	pages = {1302--17},
}

@article{zeglen_increasing_2018,
	title = {Increasing {Online} {Information} {Retention}: {Analyzing} the {Effects}},
	volume = {22},
	issn = {1179-7665},
	shorttitle = {Increasing {Online} {Information} {Retention}},
	url = {https://www.learntechlib.org/p/184660/},
	language = {en},
	number = {1},
	urldate = {2024-11-09},
	journal = {Journal of Open, Flexible, and Distance Learning},
	author = {Zeglen, Eric and Rosendale, Joseph},
	month = aug,
	year = {2018},
	note = {Publisher: Distance Education Association of New Zealand},
	pages = {22--33},
}

@article{krukar2014walk,
  title={Walk, look, remember: The influence of the gallery’s spatial layout on human memory for an art exhibition},
  author={Krukar, Jakub},
  journal={Behavioral sciences},
  volume={4},
  number={3},
  pages={181--201},
  year={2014},
  publisher={MDPI}
}

@article{krukar2015influence,
  title={The influence of an art gallery’s spatial layout on human attention to and memory of art exhibits},
  author={Krukar, Jakub},
  journal={University of Northumbria, Newcastle},
  year={2015}
}

@article{moll_optimized_2022,
	title = {Optimized virtual reality-based {Method} of {Loci} memorization techniques through increased immersion and effective memory palace designs: a feasibility study},
	volume = {27},
	shorttitle = {Optimized virtual reality-based {Method} of {Loci} memorization techniques through increased immersion and effective memory palace designs},
	url = {https://pmc.ncbi.nlm.nih.gov/articles/PMC9540171/},
	doi = {10.1007/s10055-022-00700-z},
	language = {en},
	number = {2},
	urldate = {2024-11-01},
	journal = {Virtual Reality},
	author = {Moll, Brigham and Sykes, Ed},
	month = oct,
	year = {2022},
	pmid = {36248722},
	pages = {941},
}

@inproceedings{moreno_visual_1999,
	address = {Berlin, Heidelberg},
	title = {Visual {Presentations} in {Multimedia} {Learning}: {Conditions} that {Overload} {Visual} {Working} {Memory}},
	isbn = {978-3-540-48762-3},
	shorttitle = {Visual {Presentations} in {Multimedia} {Learning}},
	doi = {10.1007/3-540-48762-X_98},
	language = {en},
	booktitle = {Visual {Information} and {Information} {Systems}},
	publisher = {Springer},
	author = {Moreno, Roxana and Mayer, Richard E.},
	editor = {Huijsmans, Dionysius P. and Smeulders, Arnold W. M.},
	year = {1999},
	pages = {798--805},
}

@article{xie_more_2017,
	title = {The more total cognitive load is reduced by cues, the better retention and transfer of multimedia learning: {A} meta-analysis and two meta-regression analyses},
	volume = {12},
	issn = {1932-6203},
	shorttitle = {The more total cognitive load is reduced by cues, the better retention and transfer of multimedia learning},
	url = {https://journals.plos.org/plosone/article?id=10.1371/journal.pone.0183884},
	doi = {10.1371/journal.pone.0183884},
	language = {en},
	number = {8},
	urldate = {2024-10-26},
	journal = {PLOS ONE},
	author = {Xie, Heping and Wang, Fuxing and Hao, Yanbin and Chen, Jiaxue and An, Jing and Wang, Yuxin and Liu, Huashan},
	month = aug,
	year = {2017},
	note = {Publisher: Public Library of Science},
	pages = {e0183884},
}

@misc{jangid_contribution_2023,
	title = {The {Contribution} of {Animation} {In} {XR} - {Chitkara} {University}},
	url = {https://www.chitkara.edu.in/blogs/the-contribution-of-animation-in-xr/},
	language = {en},
	urldate = {2024-11-01},
	journal = {Chitkara University {\textbar} Blogs},
	author = {Jangid, Sanjay},
	month = dec,
	year = {2023},
}

@inproceedings{caluya2018transferability,
  title={Transferability of spatial maps: Augmented versus virtual reality training},
  author={Caluya, Nicko R and Plopski, Alexander and Ty, Jayzon F and Sandor, Christian and Taketomi, Takafumi and Kato, Hirokazu},
  booktitle={2018 IEEE Conference on Virtual Reality and 3D User Interfaces (VR)},
  pages={387--393},
  year={2018},
  organization={IEEE}
}

@book{sweller2011cognitive,
  author    = {John Sweller and Paul Ayres and Slava Kalyuga},
  title     = {Cognitive Load Theory},
  year      = {2011},
  publisher = {Springer},
  isbn      = {978-1-4419-8126-4},
  url       = {https://doi.org/10.1007/978-1-4419-8126-4}
}

@article{sweetman2020material,
  title={Material culture, museums, and memory: Experiments in visitor recall and memory},
  author={Sweetman, Rebecca and Hadfield, Alison and O'Connor, Akira},
  journal={Visitor Studies},
  volume={23},
  number={1},
  pages={18--45},
  year={2020},
  publisher={Taylor \& Francis}
}

@article{familoni2024,
author = {Familoni, Babajide and Onyebuchi, Nneamaka},
year = {2024},
month = {04},
pages = {642-663},
title = {AUGMENTED AND VIRTUAL REALITY IN U.S. EDUCATION: A REVIEW: ANALYZING THE IMPACT, EFFECTIVENESS, AND FUTURE PROSPECTS OF AR/VR TOOLS IN ENHANCING LEARNING EXPERIENCES},
volume = {6},
journal = {International Journal of Applied Research in Social Sciences},
doi = {10.51594/ijarss.v6i4.1043}
}

@inproceedings{gavgiotaki2023gesture,
  title={Gesture-based interaction for AR systems: a short review},
  author={Gavgiotaki, Despoina and Ntoa, Stavroula and Margetis, George and Apostolakis, Konstantinos C and Stephanidis, Constantine},
  booktitle={Proceedings of the 16th International Conference on PErvasive Technologies Related to Assistive Environments},
  pages={284--292},
  year={2023}
}

@mastersthesis{cho2018spatial,
  title={How spatial presence in VR affects memory retention and motivation on second language learning: a comparison of desktop and immersive VR-based learning},
  author={Cho, Yeonhee},
  year={2018},
  school={Syracuse University}
}

@article{um2012,
author = {Um, Eunjoon and Plass, Jan and Hayward, Elizabeth and Homer, Bruce},
year = {2011},
month = {12},
pages = {485-498},
title = {Emotional Design in Multimedia Learning},
volume = {104},
journal = {Journal of Educational Psychology},
doi = {10.1037/a0026609}
}

@article{phleps_2004,
title = {Human emotion and memory: interactions of the amygdala and hippocampal complex},
journal = {Current Opinion in Neurobiology},
volume = {14},
number = {2},
pages = {198-202},
year = {2004},
issn = {0959-4388},
doi = {https://doi.org/10.1016/j.conb.2004.03.015},
url = {https://www.sciencedirect.com/science/article/pii/S0959438804000479},
author = {Elizabeth A Phelps}
}

@article{vuillemier_2005,
title = {How brains beware: neural mechanisms of emotional attention},
journal = {Trends in Cognitive Sciences},
volume = {9},
number = {12},
pages = {585-594},
year = {2005},
issn = {1364-6613},
doi = {https://doi.org/10.1016/j.tics.2005.10.011},
url = {https://www.sciencedirect.com/science/article/pii/S1364661305003025},
author = {Patrik Vuilleumier}
}

@article{schrepp2017design,
  title={Design and evaluation of a short version of the user experience questionnaire (UEQ-S)},
  author={Schrepp, Martin and Hinderks, Andreas and Thomaschewski, J{\"o}rg},
  journal={International Journal of Interactive Multimedia and Artificial Intelligence, 4 (6), 103-108.},
  year={2017},
  publisher={UNIR}
}

@article{henderson1999,
  title={High-level scene perception},
  author={Henderson, John M and Hollingworth, Andrew},
  journal={Annual review of psychology},
  volume={50},
  number={1},
  pages={243--271},
  year={1999},
  publisher={Annual Reviews 4139 El Camino Way, PO Box 10139, Palo Alto, CA 94303-0139, USA}
}

@article{rayner1992,
  title={Eye movements and scene perception},
  author={Rayner, Keith and Pollatsek, Alexander},
  journal={Canadian Journal of Psychology},
  volume={46},
  number={3},
  pages={342--376},
  year={1992},
  publisher={Canadian Psychological Association},
  doi={10.1037/h0084328}
}

@article{tversky2003,
author = {Tversky, Barbara},
year = {2003},
month = {01},
pages = {66-80},
title = {Structures Of Mental SpacesHow People Think About Space},
volume = {35},
journal = {Environment and Behavior - ENVIRON BEHAV},
doi = {10.1177/0013916502238865}
}

@article{makransky2019,
title = {Adding immersive virtual reality to a science lab simulation causes more presence but less learning},
journal = {Learning and Instruction},
volume = {60},
pages = {225-236},
year = {2019},
issn = {0959-4752},
doi = {https://doi.org/10.1016/j.learninstruc.2017.12.007},
url = {https://www.sciencedirect.com/science/article/pii/S0959475217303274},
author = {Guido Makransky and Thomas S. Terkildsen and Richard E. Mayer}
}

@inproceedings{kovacevic2022retzzles,
  title={Retzzles: Engaging Users towards Retention through Touchscreen Puzzles},
  author={Kovacevic, Nikola and Weerasinghe, Maheshya and Deja, Jordan Aiko},
  booktitle={Companion Proceedings of the 2022 Conference on Interactive Surfaces and Spaces},
  pages={14--17},
  year={2022}
}

\end{document}